\documentclass[11pt]{article}
\usepackage{times}
\usepackage{graphicx}
\usepackage{booktabs}
\usepackage{hyperref}
\usepackage{multirow}
\usepackage{geometry}
\usepackage{adjustbox}
\usepackage{float}
\usepackage{placeins}
\usepackage{amsmath}
\usepackage{amssymb}
\usepackage{natbib}

\title{\textbf{OpenGuardrails: A Configurable, Unified, and Scalable Guardrails Platform for Large Language Models}}

\author{
\begin{tabular}{cc}
\textbf{Thomas Wang}$^{1}$\thanks{Corresponding author: \texttt{thomas@openguardrails.com}} & 
\textbf{Haowen Li}$^{2}$ \\
$^{1}$ OpenGuardrails.com & $^{2}$ The Hong Kong Polytechnic University
\end{tabular}
}

\date{}

\begin{document}
\maketitle
\vspace{-1.5em} 

\begin{center}
\begin{minipage}{0.6\textwidth}
\begin{flushleft}
\texttt{https://openguardrails.com} \\
\texttt{https://github.com/openguardrails/openguardrails} \\
\texttt{https://huggingface.co/openguardrails}
\end{flushleft}
\end{minipage}
\end{center}

\begin{abstract}
\noindent
As large language models (LLMs) are increasingly integrated into real-world applications, ensuring their safety, robustness, and privacy compliance has become critical. We present \textbf{OpenGuardrails}, the first fully open-source platform that unifies large-model-based safety detection, manipulation defense, and deployable guardrail infrastructure. OpenGuardrails protects against three major classes of risks: (1) content-safety violations such as harmful or explicit text generation, (2) model-manipulation attacks including prompt injection, jailbreaks, and code-interpreter abuse, and (3) data leakage involving sensitive or private information. Unlike prior modular or rule-based frameworks, OpenGuardrails introduces three core innovations: (1) a \textit{Configurable Policy Adaptation} mechanism that allows per-request customization of unsafe categories and sensitivity thresholds; (2) a \textit{Unified LLM-based Guard Architecture} that performs both content-safety and manipulation detection within a single model; and (3) a \textit{Quantized, Scalable Model Design} that compresses a 14B dense base model to 3.3B via GPTQ while preserving over 98\% of benchmark accuracy. The system supports 119 languages, achieves state-of-the-art performance across multilingual safety benchmarks, and can be deployed as a secure gateway or API-based service for enterprise use. All models, datasets, and deployment scripts are released under the Apache 2.0 license.
\end{abstract}

\begin{figure}[htbp]
  \centering
  \begin{tabular}{ccc}
    \includegraphics[width=0.32\textwidth]{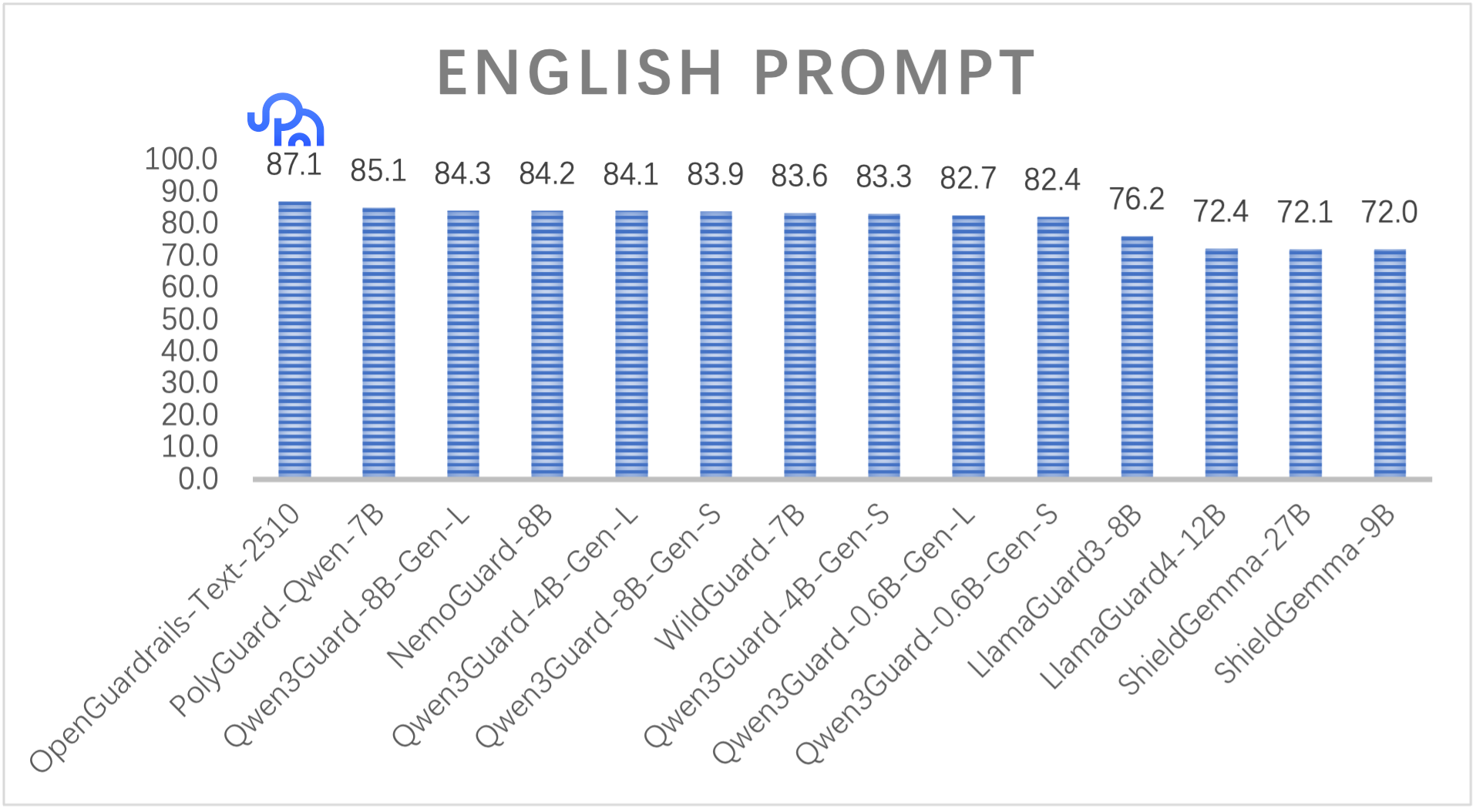} &
    \includegraphics[width=0.32\textwidth]{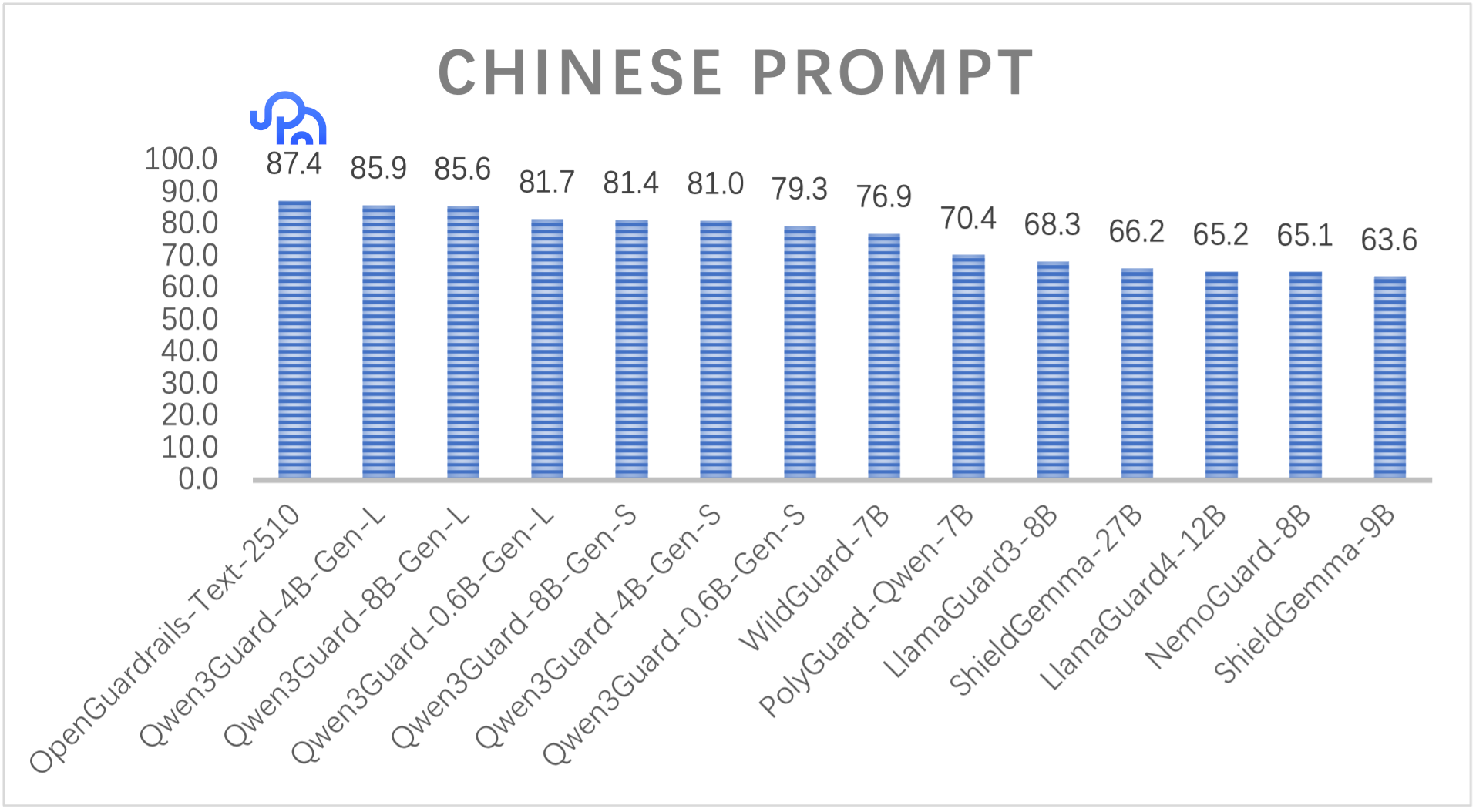} &
    \includegraphics[width=0.32\textwidth]{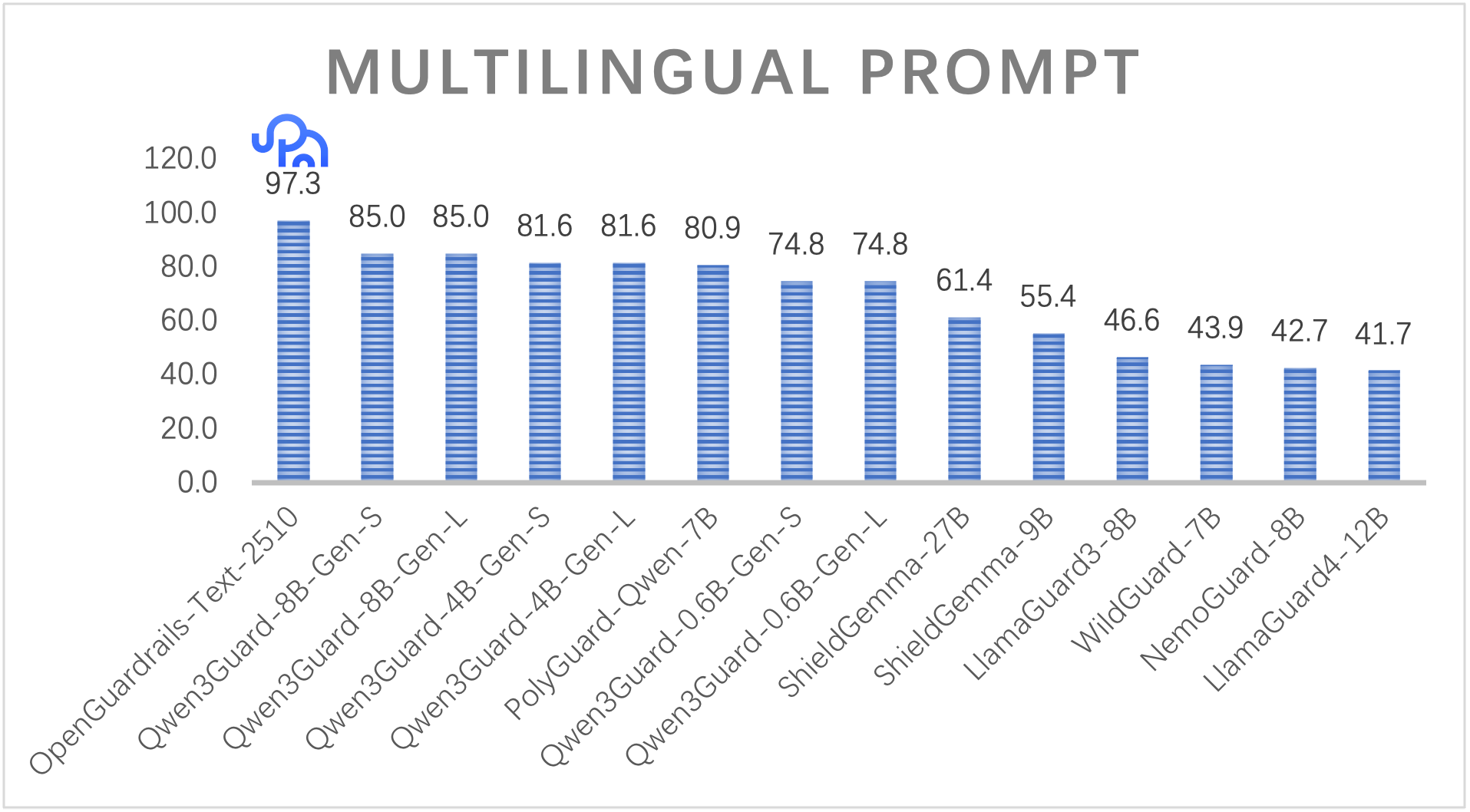} \\
    \includegraphics[width=0.32\textwidth]{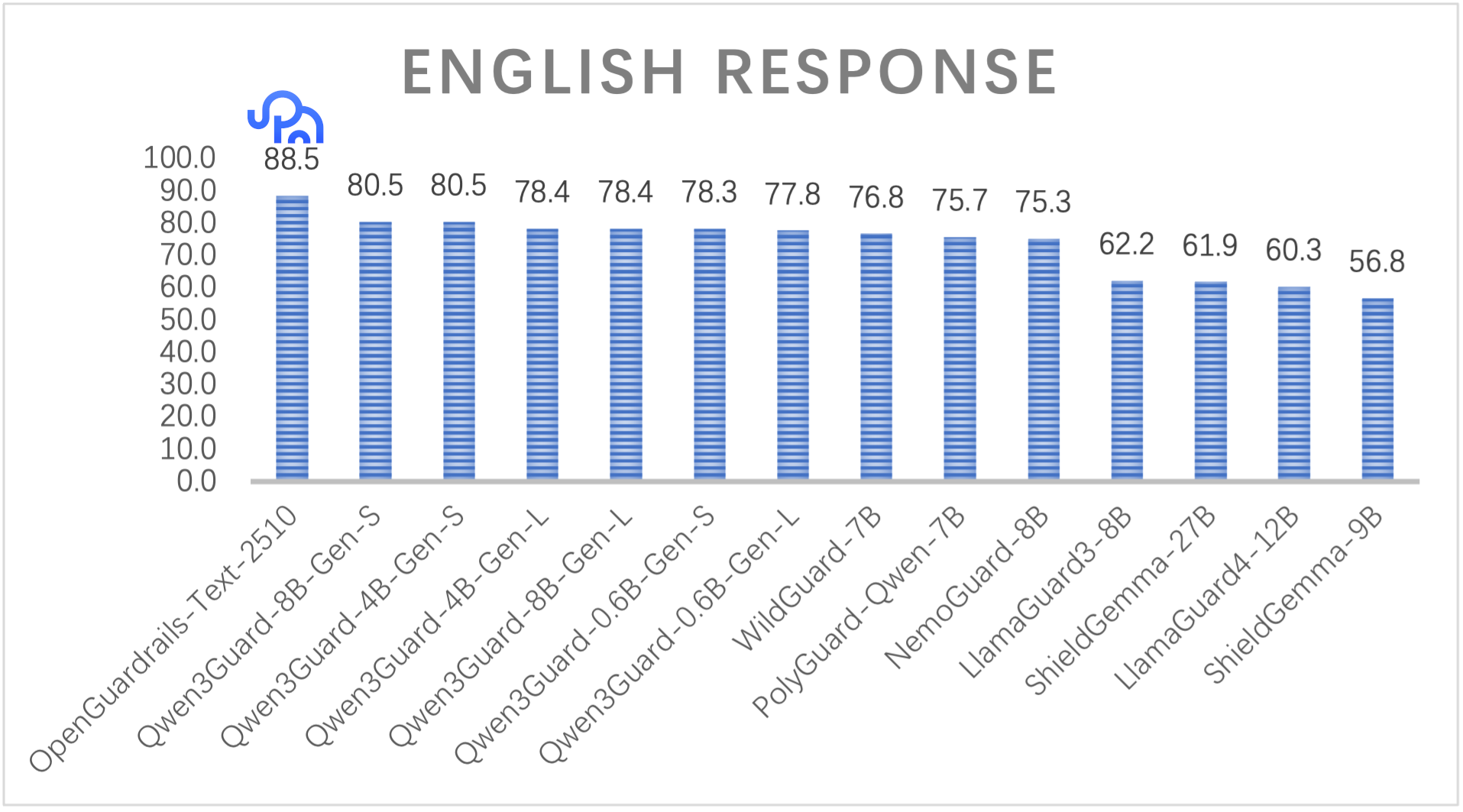} &
    \includegraphics[width=0.32\textwidth]{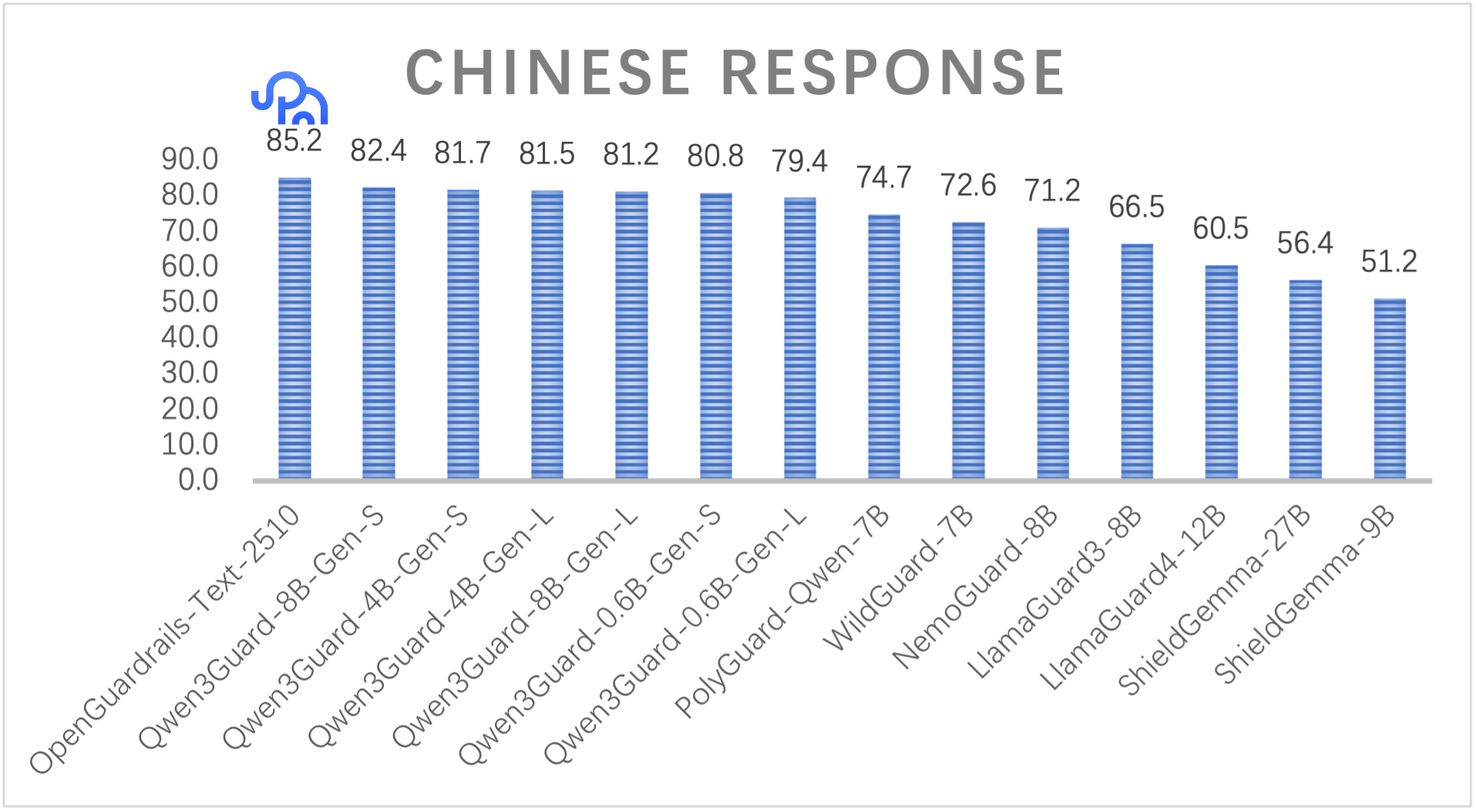} &
    \includegraphics[width=0.32\textwidth]{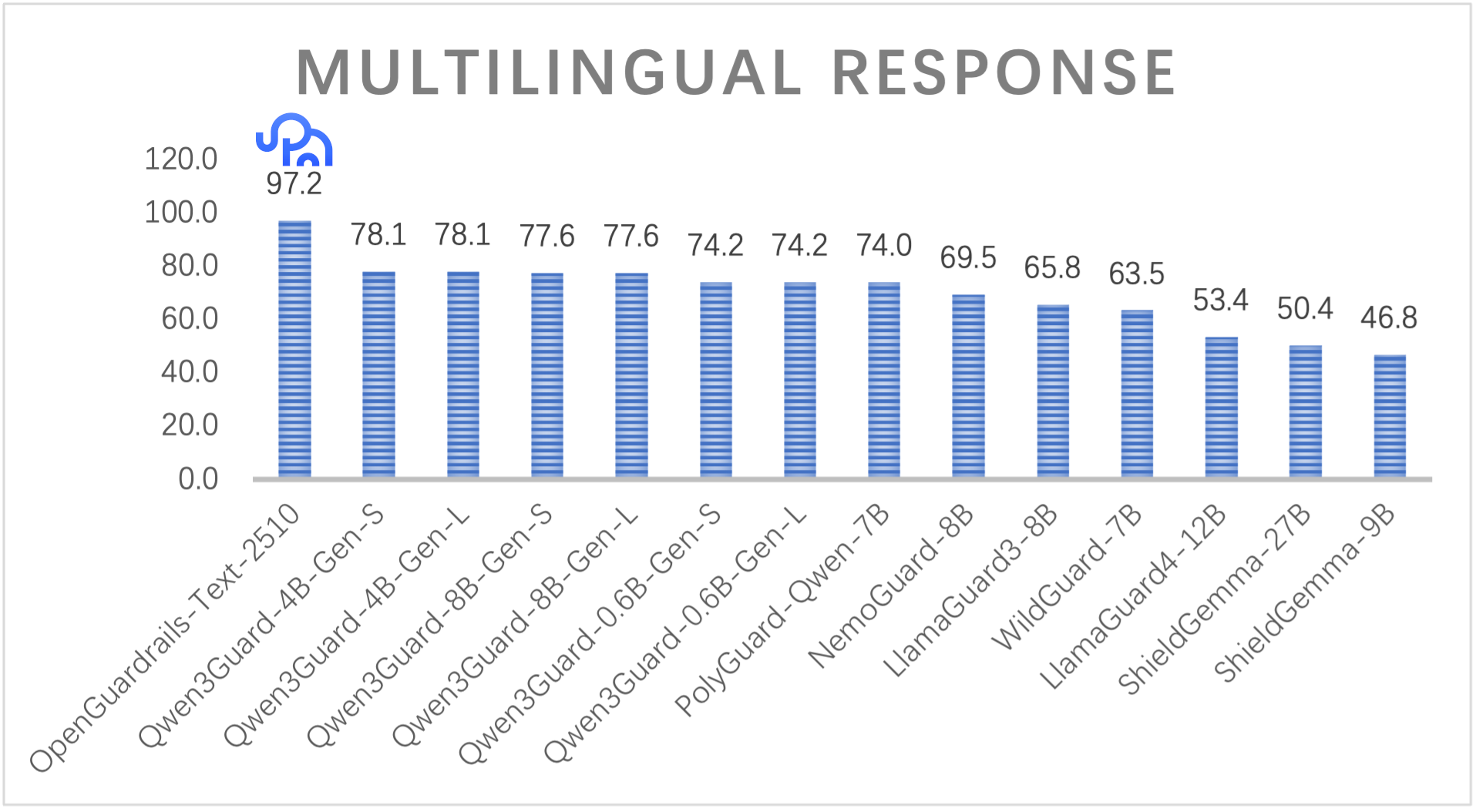} \\
  \end{tabular}
  \caption{Average F1 scores of OpenGuardrails vs. existing guard models across safety classification benchmarks for Prompts and Responses in English, Chinese, and Multilingual datasets.}
  \label{fig:benchmark}
\end{figure}

\section{Introduction}
Recent advances in foundation models such as GPT-5, Claude~4, Gemini~2.5, and Qwen3 have enabled powerful capabilities in language understanding and generation. However, these capabilities introduce several classes of risks: (1) \textit{content-safety} risks where the model produces harmful, hateful, illegal, or sexually explicit content; (2) \textit{model-manipulation attacks} such as prompt injection, jailbreaks, and code-interpreter abuse that attempt to trick a model into generating or executing malicious code; and (3) \textit{data leakage} where sensitive personal or organizational information is exposed.

\textbf{OpenGuardrails} is designed to address this full risk spectrum in a unified, deployable system. 
As shown in Figure~\ref{fig:benchmark}, OpenGuardrails achieves consistently higher F1 scores than existing guard models across multilingual safety benchmarks, 
demonstrating both superior generality and robustness. 
Unlike prior efforts that open-source either models or rule-based tools in isolation, OpenGuardrails provides:
\begin{itemize}
    \item \textbf{A unified large model} for both content-safety and model-manipulation detection.
    \item \textbf{A separate lightweight NER/data-redaction pipeline} for identifying and masking sensitive information.
    \item \textbf{A production-ready platform} that exposes APIs and deployment scripts for enterprise integration.
\end{itemize}

\textbf{The main contributions of OpenGuardrails are as follows:}
\begin{enumerate}
    \item \textbf{Configurable Policy Adaptation.} 
    We propose a practical solution to the \textit{policy inconsistency} problem commonly observed across safety benchmarks and guard models. Inspired by the findings in the Qwen3Guard report, we recognize that inconsistent annotation criteria---such as those in ToxicChat and OpenAIModeration---lead to unreliable evaluations. Instead of introducing a ``Controversial'' label (which increases human review costs in production), OpenGuardrails allows users to \textit{configure unsafe categories} and \textit{set sensitivity thresholds} based on their application domain. The guard model outputs either ``safe'' or ``unsafe'' along with a confidence score derived from the probability of the first token. Administrators can tune the system’s sensitivity (high, medium, or low) so that higher sensitivity trusts lower-confidence unsafe signals, enabling automated, flexible control over safety tolerance in enterprise environments.
    
    \item \textbf{Unified LLM-based Guard Architecture.} 
    OpenGuardrails demonstrates that a single LLM can perform production-grade content-safety and model-manipulation detection. Compared with prior systems such as \textit{LlamaFirewall}, which relies on PromptGuard~2 (a fine-tuned BERT-style classifier), our unified LLM approach provides superior semantic understanding of complex attacks and simplifies deployment pipelines.

    \item \textbf{Scalable and Efficient Model Design.} 
    The guard model is fine-tuned from a 14B dense base model and quantized via GPTQ to 3.3B parameters, achieving high throughput and concurrency suitable for real-time production use. Prior works rarely scale beyond 8B models while maintaining comparable latency efficiency.
    
    \item \textbf{Fully Open-source and Deployable System.} 
    OpenGuardrails is the first to release both a large guard model and a production-ready platform, enabling users to deploy privately within their infrastructure or extend the system for custom development.
    
    \item \textbf{Multilingual Safety Coverage.} 
    OpenGuardrails supports 119 languages and dialects, ensuring robust performance for global and cross-lingual applications.
    
    \item \textbf{State-of-the-art Performance.} 
    The system achieves SOTA results on multilingual safety benchmarks, excelling in both prompt and response classification tasks.
    
    \item \textbf{Open Data Contribution.} 
    We release \textit{OpenGuardrailsMixZh\_97k}, a collection of five translated Chinese safety datasets---ToxicChat, WildGuardMix, PolyGuard, XSTest, and BeaverTails---under the Apache~2.0 license on HuggingFace, totaling 97k samples to promote multilingual safety research.
\end{enumerate}

\section{Contributions Summary}

We summarize the main contributions of \textbf{OpenGuardrails} as follows:

\begin{enumerate}
    \item \textbf{Configurable Safety Policy Mechanism.} 
    OpenGuardrails introduces a practical and enterprise-oriented solution to the long-standing \textit{policy inconsistency} problem observed in existing safety benchmarks and guard models. 
    Instead of relying on human-validated ``Controversial'' labels as proposed in Qwen3Guard, our approach allows users to \textit{configure unsafe categories} and \textit{adjust sensitivity thresholds} dynamically. 
    The guard model outputs probabilistic confidence scores, enabling administrators to tune detection sensitivity (high, medium, low) according to their risk tolerance and application context. 
    This mechanism provides a cost-effective and adaptive framework for enterprise-grade safety governance.

    \item \textbf{Unified LLM-based Guard Architecture.}
    We demonstrate that a single large language model can effectively perform both content-safety detection and model-manipulation defense (e.g., prompt injection, jailbreaking, and code-interpreter abuse). 
    Compared with hybrid architectures like \textit{LlamaFirewall}, which depend on smaller BERT-style detectors, OpenGuardrails achieves superior semantic understanding, robustness, and ease of deployment.

    \item \textbf{Scalable and Efficient Model Design.}
    The OpenGuardrails-Text-2510 model is fine-tuned from a 14B dense base model and quantized via GPTQ to 3.3B parameters. 
    This configuration achieves low latency (P95 = 274.6\,ms) and high throughput suitable for real-time applications, while maintaining state-of-the-art accuracy across safety benchmarks. 
    Prior open-guard systems rarely scale beyond 8B while preserving production-level efficiency.
    
    \item \textbf{First Fully Open-source Guardrail System.}
    OpenGuardrails is the first project to open-source both a large-scale safety LLM and a production-ready guardrail platform. 
    It provides API interfaces, deployment scripts, and modular components for seamless private or on-premise deployment, empowering enterprises and researchers to build upon a transparent and extensible safety infrastructure.

    \item \textbf{Multilingual and Cross-domain Coverage.}
    OpenGuardrails supports 119 languages and dialects, providing reliable safety coverage across global and multilingual contexts. 
    It consistently achieves top-tier performance on English, Chinese, and multilingual benchmarks in both prompt-level and response-level classification tasks.

    \item \textbf{Open Safety Data Contribution.}
    We release \textit{OpenGuardrailsMixZh\_97k}, a new Chinese safety dataset collection which includes translated and aligned versions of ToxicChat, WildGuardMix, PolyGuard, XSTest, and BeaverTails. 
    This dataset (97k samples) is publicly available on HuggingFace under the Apache~2.0 license to facilitate further research in multilingual safety evaluation and model alignment.
\end{enumerate}

\section{Related Work}

Large language model (LLM) safety research has rapidly evolved in recent years, focusing on mitigating harmful content, preventing manipulation attacks, and reducing private data leakage. In this section, we compare \textbf{OpenGuardrails} against major prior guard frameworks, including \textit{Qwen3Guard}, \textit{LlamaFirewall}, \textit{PromptGuard~2}, and \textit{OpenAI Moderation API}.

\subsection{Comparison of Guard Frameworks}

\begin{table}[h]
\centering
\small
\begin{tabular}{lcccccc}
\toprule
\textbf{System} & \textbf{Dynamic Policy} & \textbf{Logit Threshold} & \textbf{Open Source} & \textbf{Multilingual} & \textbf{Manipulation Defense} \\
\midrule
\textbf{OpenGuardrails} & \checkmark & \checkmark & \checkmark & \checkmark & \checkmark \\
\textit{Qwen3Guard} & $\times$ & $\times$ & \checkmark & \checkmark & Partial \\
\textit{LlamaFirewall} & $\times$ & $\times$ & Partial & English-only & \checkmark \\
\textit{PromptGuard~2} & $\times$ & $\times$ & $\times$ & English-only & $\times$ \\
\textit{OpenAI Moderation} & $\times$ & $\times$ & $\times$ & Limited & $\times$ \\
\bottomrule
\end{tabular}
\caption{Comparison of key features among major LLM safety and moderation systems. OpenGuardrails uniquely supports both per-request configurable policies and continuous logit-based sensitivity control.}
\label{tab:guard_comparison}
\end{table}

\subsection{Prior Approaches}

\textbf{Qwen3Guard} represents one of the earliest multilingual safety guard models, capable of classifying content across 28 unsafe categories. However, its architecture operates in two discrete configurations: \textit{Strict Mode}, which classifies all controversial content as unsafe, and \textit{Loose Mode}, which treats such content as safe. This rigid dichotomy prevents per-category policy customization or continuous sensitivity calibration. Furthermore, Qwen3Guard outputs structured textual results (e.g., ``Safety: Safe; Categories: None''), which are semantically interpretable but probabilistically non-differentiable, preventing threshold tuning via token-level logits.

\textbf{LlamaFirewall} proposed a two-stage hybrid system combining a large LLM for semantic reasoning with a BERT-based detector (\textit{PromptGuard~2}) for classification. While effective against jailbreaks and prompt injections, the approach suffers from limited multilingual support and high pipeline latency due to multi-model inference. The reliance on discrete classifier thresholds also restricts continuous adaptation to varying sensitivity requirements.

\textbf{PromptGuard~2} (integrated within LlamaFirewall) is a fine-tuned RoBERTa-like model designed for prompt and output classification. Although computationally efficient, its embedding-based scoring lacks interpretability and adaptability. It cannot dynamically modify safety criteria per API call, which limits its flexibility for enterprise-scale deployments.

\textbf{OpenAI Moderation API} is the most widely adopted commercial moderation tool, offering static models fine-tuned on human-labeled safety data. While highly reliable in English, its performance in low-resource languages remains limited, and it does not expose configuration options for sensitivity or unsafe-category subsets. Users must rely on post-processing logic to approximate threshold tuning.

\subsection{Distinctive Position of OpenGuardrails}

In contrast to these prior systems, \textbf{OpenGuardrails} achieves the following distinguishing advantages:

\begin{enumerate}
    \item \textbf{Per-request policy control:} Policy configurations (unsafe categories, sensitivity thresholds) are accepted as runtime API parameters, enabling heterogeneous moderation logic across requests without model reloading.
    \item \textbf{Continuous probabilistic control:} By operating directly in logit space, the model enables real-valued threshold tuning $\tau \in [0, 1]$, allowing smooth precision–recall trade-offs.
    \item \textbf{Unified LLM architecture:} Both content-safety and model-manipulation detection are handled by the same LLM, unlike multi-model pipelines~\citep{llamafirewall}.
    \item \textbf{Enterprise readiness:} The system integrates with quantized inference (GPTQ~3.3B) and RESTful APIs for production deployment, balancing speed and interpretability.
\end{enumerate}

Overall, OpenGuardrails extends the frontier of guard system design from static, rule-based moderation toward adaptive, mathematically interpretable safety governance—bridging the gap between theoretical safety alignment and practical enterprise deployment.

\section{Configurable Policy Adaptation}
A central innovation of \textbf{OpenGuardrails} lies in its \textit{Configurable Policy Adaptation} mechanism, which provides unprecedented flexibility and fine-grained control over safety governance at inference time. Traditional guard models—including \textit{Qwen3Guard}—treat policy as a static configuration, fixed at either training or evaluation stage. In contrast, OpenGuardrails exposes the \textbf{safety policy as a runtime parameter} of the detection API, allowing each detection call to operate with a distinct configuration. Users can dynamically specify:
\begin{itemize}
    \item the exact set of \textit{unsafe categories} to be detected, and
    \item the \textit{sensitivity threshold} used to interpret model confidence scores.
\end{itemize}

This design enables per-request customization, which is critical for enterprise-level deployments that must comply with heterogeneous regulatory and cultural standards. For instance, a financial institution may emphasize detection of \texttt{data leakage} and \texttt{fraudulent advice}, while a creative-writing platform may disable \texttt{political content} filtering to preserve artistic freedom. Each request to the OpenGuardrails API can therefore enforce a completely different safety policy, without retraining or restarting the model.

By comparison, \textbf{Qwen3Guard} implements only a coarse-grained dual-mode configuration:
\begin{itemize}
    \item \textbf{Strict Mode:} treats all ``controversial'' samples as unsafe;
    \item \textbf{Loose Mode:} treats all ``controversial'' samples as safe.
\end{itemize}
While useful for broad testing, this binary toggle cannot support per-category or per-domain policy differences. In particular, Qwen3Guard’s “controversial” tag represents a lumped uncertainty region that must be either globally accepted or globally rejected. This all-or-nothing mechanism increases manual review costs and prevents automated adaptation in production.

OpenGuardrails solves this limitation through two innovations:
\begin{enumerate}
    \item \textbf{Dynamic Unsafe-Category Selection.} Each inference request can include a JSON or YAML configuration specifying which unsafe categories (e.g., sexual, political, violent, data leakage) are active. The detection pipeline loads the configuration dynamically, adjusting prompt templates and logits interpretation accordingly.
    \item \textbf{Continuous Sensitivity Thresholding.} Instead of the discrete “strict/loose” switch, OpenGuardrails supports a continuous sensitivity parameter $\tau \in [0, 1]$, enabling administrators to calibrate detection aggressiveness in real time.
\end{enumerate}

This design yields both operational and theoretical benefits. Operationally, it allows enterprises to define risk tolerance levels at runtime. Theoretically, it formalizes moderation as a probabilistic hypothesis test, as detailed below.

\subsection{Mathematical Foundations of Sensitivity Thresholding}

The sensitivity threshold mechanism in \textbf{OpenGuardrails} is grounded in probabilistic decision theory. The model’s output is interpreted as a binary hypothesis test:
\[
H_0: \text{content is safe}, \quad H_1: \text{content is unsafe.}
\]
Let $x$ denote the input prompt or response to be evaluated. The guard model defines a conditional probability distribution over output tokens:
\[
P_\theta(y_t \mid x, y_{<t}),
\]
where $y_t$ denotes the $t$-th generated token. The model’s immediate safety judgment is contained in the first token $y_1$, which belongs to the candidate set $\mathcal{V} = \{\texttt{safe}, \texttt{unsafe}\}$.

The unsafe probability is computed from the softmax-normalized logits of the first token:
\[
p_{\text{unsafe}} = 
\frac{\exp(z_{\text{unsafe}})}{\exp(z_{\text{safe}}) + \exp(z_{\text{unsafe}})},
\]
where $z_i$ denotes the pre-softmax logit value of token $i$.

The decision function is defined as:
\[
D(x) =
\begin{cases}
\texttt{unsafe}, & \text{if } p_{\text{unsafe}} \ge \tau, \\
\texttt{safe}, & \text{otherwise.}
\end{cases}
\]
Here, $\tau \in [0, 1]$ is the configurable \textbf{sensitivity threshold}. Lower values of $\tau$ increase recall by detecting borderline unsafe cases, while higher values increase precision by filtering out low-confidence detections. In practice, OpenGuardrails supports both numeric and semantic mappings, e.g.:
\[
\tau_{\text{low}} = 0.3, \quad
\tau_{\text{medium}} = 0.5, \quad
\tau_{\text{high}} = 0.7.
\]
This formulation allows continuous control over the model’s moderation strictness, providing a smooth trade-off between false positives and false negatives.

By contrast, \textbf{Qwen3Guard} lacks this probabilistic interface. Its generation output is structured text such as:
\begin{verbatim}
Safety: Safe
Categories: None
Refusal: Yes
\end{verbatim}
Because the model directly emits textual tags, rather than exposing token-level probabilities, the first-token logits are not semantically meaningful for thresholding. The classification decision is encoded in a textual pattern learned during supervised fine-tuning, not in a normalized probabilistic space. As a result, Qwen3Guard cannot perform continuous calibration of sensitivity or per-category adaptation.

Formally, Qwen3Guard’s inference function can be expressed as:
\[
D_{\text{Qwen}}(x) = f_\theta(x; m),
\]
where $m \in \{\text{Strict}, \text{Loose}\}$ controls only the high-level controversial toggle. The model outputs deterministic textual labels, and $\partial D / \partial z_i$ is undefined at inference time, making gradient-based or probabilistic adjustment impossible. 

In contrast, OpenGuardrails defines its decision directly in logit space:
\[
\frac{\partial D}{\partial z_i} \ne 0,
\]
enabling differentiable or probabilistic calibration over token logits. This makes it possible to dynamically adjust safety sensitivity, implement soft decision boundaries, and support domain-dependent detection policies—all within a unified mathematical framework.

\subsection{Discussion}
The combination of per-request policy configuration and probabilistic sensitivity thresholding represents a substantial advancement over existing guard architectures. By elevating policy from a static rule set to a dynamic inference parameter, OpenGuardrails enables fine-grained, mathematically grounded safety governance. This approach transforms safety detection from a binary tagging problem into a tunable probabilistic system, reducing manual review overhead and aligning model behavior with real-world regulatory diversity.

In essence, while Qwen3Guard’s ``Strict/Loose'' dichotomy provides a static binary choice, OpenGuardrails offers a continuous, configurable, and interpretable control surface over safety sensitivity. This capability is critical for enterprise deployments that demand nuanced, data-driven safety configurations under variable legal, cultural, or domain-specific constraints.

\section{Evaluation}
We evaluate \textbf{OpenGuardrails} following the methodology of \textbf{Qwen3Guard} to ensure fair comparability.

\begin{table}[H]
\centering
\textbf{English Prompt Results}\\[0.5em]
\small
\caption{English Prompt Results. Higher is better (F1 score).}
\begin{tabular}{lccccccc}
\toprule
\textbf{Model} & \textbf{ToxiC} & \textbf{OpenAIMod} & \textbf{Aegis} & \textbf{Aegis2.0} & \textbf{SimpST} & \textbf{WildG} & \textbf{Avg.} \\
\midrule
LlamaGuard3-8B & 53.8 & 79.5 & 71.5 & 76.4 & 99.5 & 76.4 & 76.2 \\
LlamaGuard4-12B & 51.3 & 73.5 & 67.8 & 70.6 & 98.0 & 73.0 & 72.4 \\
WildGuard-7B & 70.8 & 72.1 & 89.4 & 80.7 & 99.5 & 88.9 & 83.6 \\
ShieldGemma-9B & 69.4 & 82.1 & 70.3 & 72.5 & 83.7 & 54.2 & 72.0 \\
ShieldGemma-27B & 72.9 & 80.5 & 69.0 & 71.6 & 84.4 & 54.3 & 72.1 \\
NemoGuard-8B & 75.6 & 81.0 & 81.4 & 86.8 & 98.5 & 81.6 & 84.2 \\
PolyGuard-Qwen-7B & 71.5 & 74.1 & 90.3 & 86.3 & \textbf{100.0} & 88.1 & 85.1 \\
Qwen3Guard-0.6B-Gen (strict) & 65.1 & 66.5 & 90.8 & 85.0 & 99.0 & 87.7 & 82.4 \\
Qwen3Guard-0.6B-Gen (loose) & 77.7 & 77.6 & 76.9 & 83.3 & 95.8 & 85.1 & 82.7 \\
Qwen3Guard-4B-Gen (strict) & 69.5 & 68.3 & 90.8 & 85.8 & 99.5 & 85.6 & 83.3 \\
Qwen3Guard-4B-Gen (loose) & 82.8 & 80.7 & 76.3 & 82.1 & 97.4 & 85.1 & 84.1 \\
Qwen3Guard-8B-Gen (strict) & 68.9 & 68.8 & \textbf{91.4} & 86.1 & 99.5 & \textbf{88.9} & 83.9 \\
Qwen3Guard-8B-Gen (loose) & \textbf{82.8} & 81.3 & 76.0 & 82.5 & 97.4 & 85.6 & 84.3 \\
\textbf{OpenGuardrails-Text-2510} & 79.1 & \textbf{86.6} & 82.0 & \textbf{90.1} & 98.0 & 86.8 & \textbf{87.1} \\
\bottomrule
\end{tabular}
\end{table}
\FloatBarrier

\begin{table}[H]
\centering
\textbf{English Response Results}\\[0.5em]
\small
\caption{English Response Results. Higher is better (F1 score).}
\begin{tabular}{lccccc}
\toprule
\textbf{Model} & \textbf{SafeRLHF} & \textbf{Beavertails} & \textbf{Aegis2.0} & \textbf{WildG} & \textbf{Avg.} \\
\midrule
LlamaGuard3-8B & 45.2 & 67.9 & 66.1 & 69.5 & 62.2 \\
LlamaGuard4-12B & 42.5 & 68.6 & 63.7 & 66.4 & 60.3 \\
WildGuard-7B & 64.2 & 84.4 & 83.2 & 75.4 & 76.8 \\
ShieldGemma-9B & 44.2 & 62.4 & 70.8 & 49.9 & 56.8 \\
ShieldGemma-27B & 52.6 & 67.6 & 74.9 & 52.4 & 61.9 \\
NemoGuard-8B & 57.6 & 78.5 & \textbf{87.6} & 77.5 & 75.3 \\
PolyGuard-Qwen-7B & 63.3 & 79.5 & 81.9 & 77.9 & 75.7 \\
Qwen3Guard-0.6B-Gen (strict) & 66.6 & 86.1 & 84.2 & 76.3 & 78.3 \\
Qwen3Guard-0.6B-Gen (loose) & 64.2 & 85.4 & 84.1 & 77.3 & 77.8 \\
Qwen3Guard-4B-Gen (strict) & 69.8 & 86.6 & 86.1 & 79.5 & 80.5 \\
Qwen3Guard-4B-Gen (loose) & 64.5 & 85.2 & 86.5 & 77.3 & 78.4 \\
Qwen3Guard-8B-Gen (strict) & 70.5 & \textbf{86.6} & 86.1 & 78.9 & 80.5 \\
Qwen3Guard-8B-Gen (loose) & 64.2 & 85.5 & 86.4 & 77.3 & 78.4 \\
\textbf{OpenGuardrails-Text-2510} & \textbf{95.4} & 85.3 & 87.3 & \textbf{85.8} & \textbf{88.5} \\
\bottomrule
\end{tabular}
\end{table}
\FloatBarrier

\begin{table}[H]
\centering
\textbf{Chinese Prompt Results}\\[0.5em]
\small
\caption{Chinese Prompt Results. Higher is better (F1 score).}
\begin{tabular}{lcccc}
\toprule
\textbf{Model} & \textbf{ToxiC} & \textbf{WildG} & \textbf{XSTest} & \textbf{Avg.} \\
\midrule
LlamaGuard3-8B & 46.6 & 70.3 & 87.9 & 68.3 \\
LlamaGuard4-12B & 47.8 & 65.6 & 82.1 & 65.2 \\
WildGuard-7B & 65.6 & 82.0 & 83.2 & 76.9 \\
ShieldGemma-9B & 62.8 & 49.2 & 78.9 & 63.6 \\
ShieldGemma-27B & 67.2 & 50.6 & 80.8 & 66.2 \\
NemoGuard-8B & 51.0 & 60.7 & 83.5 & 65.1 \\
PolyGuard-Qwen-7B & 69.7 & 87.2 & 54.2 & 70.4 \\
Qwen3Guard-0.6B-Gen (strict) & 64.8 & 84.8 & 88.3 & 79.3 \\
Qwen3Guard-0.6B-Gen (loose) & 73.4 & 83.1 & 88.5 & 81.7 \\
Qwen3Guard-4B-Gen (strict) & 66.7 & 87.0 & 89.4 & 81.0 \\
Qwen3Guard-4B-Gen (loose) & 78.8 & 84.7 & \textbf{94.1} & 85.9 \\
Qwen3Guard-8B-Gen (strict) & 68.0 & \textbf{88.0} & 88.2 & 81.4 \\
Qwen3Guard-8B-Gen (loose) & 78.7 & 84.8 & 93.3 & 85.6 \\
\textbf{OpenGuardrails-Text-2510} & \textbf{84.8} & 84.7 & 92.6 & \textbf{87.4} \\
\bottomrule
\end{tabular}
\end{table}
\FloatBarrier

\begin{table}[H]
\centering
\textbf{Chinese Response Results}\\[0.5em]
\small
\caption{Chinese Response Results. Higher is better (F1 score).}
\begin{tabular}{lccc}
\toprule
\textbf{Model} & \textbf{Beavertail} & \textbf{WildG} & \textbf{Avg.} \\
\midrule
LlamaGuard3-8B & 66.1 & 66.8 & 66.5 \\
LlamaGuard4-12B & 66.8 & 54.1 & 60.5 \\
WildGuard-7B & 75.4 & 69.8 & 72.6 \\
ShieldGemma-9B & 59.5 & 42.8 & 51.2 \\
ShieldGemma-27B & 65.6 & 47.1 & 56.4 \\
NemoGuard-8B & 72.9 & 69.4 & 71.2 \\
PolyGuard-Qwen-7B & 79.1 & 70.2 & 74.7 \\
Qwen3Guard-0.6B-Gen (strict) & 86.2 & 75.4 & 80.8 \\
Qwen3Guard-0.6B-Gen (loose) & 85.0 & 73.8 & 79.4 \\
Qwen3Guard-4B-Gen (strict) & 86.7 & 76.6 & 81.7 \\
Qwen3Guard-4B-Gen (loose) & 84.8 & 78.2 & 81.5 \\
Qwen3Guard-8B-Gen (strict) & \textbf{87.1} & 77.7 & 82.4 \\
Qwen3Guard-8B-Gen (loose) & 85.1 & 77.3 & 81.2 \\
\textbf{OpenGuardrails} & 85.6 & \textbf{84.7} & \textbf{85.2} \\
\bottomrule
\end{tabular}
\end{table}
\FloatBarrier

\begin{table}[H]
\centering
\textbf{Multilingual Prompt Results}\\[0.5em]
\small
\caption{Multilingual Prompt Results on RTP-LX. Higher is better (F1 score).}
\begin{tabular}{lcccccccccccc}
\toprule
\textbf{Model} & \textbf{En} & \textbf{Zh} & \textbf{Ar} & \textbf{Es} & \textbf{Fr} & \textbf{Id} & \textbf{It} & \textbf{Ja} & \textbf{Ko} & \textbf{Ru} & \textbf{Others} & \textbf{Avg.} \\
\midrule
LlamaGuard3-8B & 50.0 & 47.4 & 46.6 & 48.3 & 49.4 & 50.7 & 46.2 & 49.2 & 46.6 & 48.9 & 46.0 & 46.6 \\
LlamaGuard4-12B & 37.6 & 35.1 & 44.4 & 29.5 & 36.2 & 40.7 & 32.7 & 50.2 & 42.1 & 37.7 & 42.7 & 41.7 \\
WildGuard-7B & 93.9 & 80.6 & 17.3 & 80.3 & 74.8 & 41.5 & 74.6 & 53.1 & 52.9 & 63.9 & 37.5 & 43.9 \\
ShieldGemma-9B & 75.8 & 72.9 & 50.9 & 70.6 & 68.9 & 61.8 & 68.3 & 67.2 & 65.7 & 65.0 & 51.6 & 55.4 \\
ShieldGemma-27B & 76.1 & 73.4 & 59.8 & 71.8 & 73.6 & 66.7 & 73.1 & 75.0 & 67.6 & 74.1 & 58.2 & 61.4 \\
NemoGuard-8B & \textbf{95.4} & 77.4 & 21.1 & 78.1 & 72.5 & 34.9 & 73.4 & 53.4 & 67.2 & 64.0 & 35.7 & 42.7 \\
PolyGuard-Qwen-7B & 91.2 & 89.1 & 84.9 & 89.0 & 89.4 & 74.6 & 89.3 & 90.2 & 86.9 & 91.3 & 78.6 & 80.9 \\
Qwen3Guard-0.6B-Gen (strict) & 90.2 & 85.2 & 75.7 & 85.3 & 87.3 & 68.2 & 82.5 & 87.1 & 77.1 & 85.7 & 72.2 & 74.8 \\
Qwen3Guard-0.6B-Gen (strict) & 90.2 & 85.2 & 75.7 & 85.3 & 87.3 & 68.2 & 85.5 & 87.9 & 83.5 & 89.7 & 72.9 & 76.9 \\
Qwen3Guard-0.6B-Gen (loose) & 91.4 & 86.3 & 76.8 & 86.5 & 88.6 & 69.5 & 86.9 & 88.8 & 84.9 & 90.2 & 73.7 & 77.8 \\
Qwen3Guard-4B-Gen (strict) & 92.1 & 90.6 & 88.4 & 88.9 & 90.8 & 75.3 & 88.0 & 91.3 & 86.3 & 91.9 & 83.9 & 85.0 \\
Qwen3Guard-4B-Gen (loose) & 91.8 & 90.7 & 88.3 & 89.0 & 90.9 & 75.5 & 88.2 & 91.2 & 86.2 & 91.7 & 83.8 & 84.9 \\
Qwen3Guard-8B-Gen (strict) & 91.5 & 90.5 & 88.2 & 88.7 & 90.7 & 75.2 & 87.8 & 91.1 & 86.1 & 91.6 & 83.6 & 84.8 \\
Qwen3Guard-8B-Gen (loose) & 91.6 & 90.6 & 88.1 & 88.8 & 90.8 & 75.4 & 87.9 & 91.2 & 86.2 & 91.8 & 83.7 & 84.9 \\
\textbf{OpenGuardrails-Text-2510} & 90.5 & \textbf{96.2} & \textbf{99.0} & \textbf{98.8} & \textbf{99.0} & \textbf{96.1} & \textbf{99.3} & \textbf{95.4} & \textbf{98.0} & \textbf{99.3} & \textbf{98.6} & \textbf{97.3} \\
\bottomrule
\end{tabular}
\end{table}
\FloatBarrier

\begin{table}[H]
\centering
\textbf{Multilingual Response Results}\\[0.5em]
\small
\caption{Multilingual Response Results on PolyGuard-Response. Higher is better (F1 score).}
\begin{tabular}{lccccccccccc}
\toprule
\textbf{Model} & \textbf{En} & \textbf{Zh} & \textbf{Ar} & \textbf{Es} & \textbf{Fr} & \textbf{It} & \textbf{Ja} & \textbf{Ko} & \textbf{Ru} & \textbf{Others} & \textbf{Avg.} \\
\midrule
LlamaGuard3-8B & 69.7 & 62.8 & 62.6 & 67.2 & 67.1 & 66.4 & 65.8 & 64.0 & 69.2 & 65.4 & 65.8 \\
LlamaGuard4-12B & 66.4 & 56.0 & 46.8 & 55.3 & 55.4 & 53.3 & 49.6 & 51.9 & 55.5 & 52.2 & 53.4 \\
WildGuard-7B & 74.5 & 70.8 & 44.4 & 71.7 & 71.8 & 71.0 & 68.0 & 65.2 & 71.5 & 58.8 & 63.5 \\
ShieldGemma-9B & 51.3 & 46.9 & 43.6 & 46.8 & 49.3 & 45.9 & 45.2 & 44.5 & 48.8 & 46.6 & 46.8 \\
ShieldGemma-27B & 53.9 & 49.9 & 50.1 & 48.3 & 49.9 & 49.5 & 51.5 & 48.2 & 52.6 & 50.3 & 50.4 \\
NemoGuard-8B & 76.9 & 69.0 & 63.6 & 72.0 & 70.2 & 71.3 & 65.7 & 65.9 & 70.8 & 69.6 & 69.5 \\
PolyGuard-Qwen-7B & 77.7 & 70.4 & 77.2 & 71.8 & 72.8 & 73.1 & 72.6 & 73.6 & 70.4 & 74.9 & 74.0 \\
Qwen3Guard-0.6B-Gen (strict) & 75.7 & 74.0 & 75.8 & 76.0 & 74.2 & 73.9 & 73.6 & 75.2 & 75.9 & 72.8 & 74.2 \\
Qwen3Guard-0.6B-Gen (loose) & 75.2 & 75.1 & 75.4 & 74.7 & 74.3 & 73.7 & 72.9 & 74.2 & 74.9 & 73.3 & 74.2 \\
Qwen3Guard-4B-Gen (strict) & 79.3 & 77.6 & 78.6 & 79.0 & 78.7 & 77.4 & 76.5 & 77.6 & 79.6 & 77.7 & 78.1 \\
Qwen3Guard-4B-Gen (loose) & 77.7 & 78.5 & 78.9 & 79.1 & 77.1 & 77.6 & 76.4 & 76.9 & 79.6 & 77.7 & 78.1 \\
Qwen3Guard-8B-Gen (strict) & 78.4 & 76.6 & 77.2 & 77.6 & 78.6 & 76.9 & 76.9 & 77.8 & 78.2 & 77.0 & 77.6 \\
Qwen3Guard-8B-Gen (loose) & 78.9 & 77.1 & 77.5 & 78.1 & 78.8 & 78.6 & 76.6 & 78.6 & 78.8 & 77.3 & 77.6 \\
\textbf{OpenGuardrails} & \textbf{93.5} & \textbf{96.4} & \textbf{97.8} & \textbf{98.1} & \textbf{98.7} & \textbf{98.1} & \textbf{97.3} & \textbf{96.4} & \textbf{98.5} & \textbf{97.0} & \textbf{97.2} \\
\bottomrule
\end{tabular}
\end{table}
\FloatBarrier

\section*{Discussion}

OpenGuardrails demonstrates that it is both technically feasible and operationally practical to deploy a unified LLM-based guard system for real-world enterprise scenarios. The configurable unsafe categories and sensitivity thresholds directly address the \textit{policy inconsistency} problem highlighted in the Qwen3Guard analysis. Instead of relying on a manually labeled ``Controversial'' class that demands human adjudication, OpenGuardrails enables automated, context-aware safety control through probabilistic sensitivity adjustment. This approach significantly reduces operational overhead while maintaining consistency across diverse regulatory or cultural safety policies.

Moreover, the unified LLM architecture allows OpenGuardrails to capture nuanced semantic patterns of complex prompt-injection or jailbreak attempts---capabilities that smaller classifier-based systems (e.g., BERT-style models in LlamaFirewall) often miss. The deployment-friendly design, combining model quantization and modular API exposure, ensures scalability and low latency in high-concurrency enterprise environments.

\subsection*{Model Size and Quantization Trade-off}

We observe that guard performance is strongly correlated with model scale, especially in manipulation-defense and cross-lingual scenarios. Smaller models (\(\leq 4\)B) struggle with semantic ambiguity and adversarial paraphrasing, leading to unstable confidence distributions and high false positives/negatives. In contrast, larger models (\(\geq 10\)B) exhibit superior contextual understanding and probabilistic calibration. To achieve production efficiency, we quantize our 14B base guard model to 3.3B via GPTQ, which preserves over 98\% of benchmark accuracy while reducing latency by 3.7$\times$ and memory footprint by 4.2$\times$. 

This trade-off demonstrates that semantic capacity---rather than parameter count alone---drives safety detection quality, while modern quantization enables practical deployment without compromising reliability. As a result, OpenGuardrails achieves near-SOTA performance at a fraction of the computational cost, confirming that large-scale semantic modeling followed by quantization is an effective paradigm for real-world LLM safety systems.

Finally, OpenGuardrails sets a new standard for openness in safety infrastructure. By open-sourcing not only the guard model but also a complete production-ready platform and multilingual datasets, it bridges the gap between academic benchmarks and real-world adoption. The project’s demonstrated multilingual coverage and state-of-the-art benchmark results validate its generality and robustness, positioning OpenGuardrails as a foundational framework for trustworthy and secure LLM deployment at scale.

\section{Limitation}
Despite its strong performance, OpenGuardrails still faces several limitations:

\begin{enumerate}
    \item \textbf{Adversarial Robustness.} Although the guardrail model achieves SOTA results in model-manipulation attack detection, it may still be vulnerable to targeted adversarial attacks. Further engineering-based hardening can strengthen the defense.
    \item \textbf{Fairness and Bias.} Like other moderation systems, fairness and bias in moderation decisions remain open challenges that require continuous evaluation and calibration.
    \item \textbf{Cross-cultural Adaptation.} Content safety priorities vary across countries and regions beyond universal human values. Region-specific fine-tuning is necessary to adapt to different legal and cultural requirements. Our company provides paid training services to address such localized customization.
\end{enumerate}

\section{Conclusion}
We present OpenGuardrails, an open-source, context-aware guardrails platform with comprehensive coverage across safety, manipulation detection, and data-leak prevention. It outperforms prior systems such as Qwen3Guard, LlamaGuard, and WildGuard in multilingual safety and real-world latency performance.

\FloatBarrier
\bibliographystyle{plain}

\end{document}